\documentclass{article}
\usepackage{main}
\usepackage{amsmath,amssymb}
\usepackage{graphicx}
\usepackage{natbib}
\usepackage{caption}
\usepackage{newfloat}
\usepackage{listings}
\usepackage{xcolor}
\usepackage{multirow}
\usepackage{multicol}
\usepackage{booktabs}
\usepackage{arydshln}
\usepackage{color}
\usepackage{makecell}
\usepackage{subcaption}
\usepackage{tabularx}
\usepackage{colortbl}
\usepackage{pifont}
\usepackage{wrapfig}

\definecolor{lightgreen}{RGB}{150, 250, 170}
\definecolor{lightred}{RGB}{255, 100, 100}
\definecolor{yuangreen}{HTML}{00CC70}
\hypersetup{allcolors=yuangreen}
\usepackage[most]{tcolorbox}
\usepackage{float}
\usepackage{xspace}
\usepackage{enumitem}
\tcbset{
  promptbox/.style={
    width=0.95\textwidth,
    colback=blue!5!gray!5,
    colframe=blue!30!gray,
    colbacktitle=blue!5!gray!5!,
    fonttitle=\bfseries\color{black},
    coltitle=black,
    boxrule=0.3mm,
    arc=3mm,
    top=6pt,
    bottom=6pt,
    left=10pt,
    right=10pt,
    enhanced,
    center,
    attach boxed title to top left={xshift=5mm, yshift=-3mm},
    boxed title style={
      arc=2mm,
      colframe=blue!30!gray,
      colback=blue!15!gray!15!
    },
    title filled,
    drop shadow={xshift=0.5mm, yshift=-0.5mm, fill=gray!30}
  }
}

\newtcolorbox{PromptBox}[2][]{promptbox, title=#2, #1}

\newcommand{\ours}{\textsc{GistRank}}

\newcommand{\secref}[1]{Section~\ref{#1}}

\newcommand{\figref}[1]{Figure~\ref{#1}}
 
\newcommand{\equref}[1]{Equation~\ref{#1}}

\title{\ours: Gist-Compressed Listwise Ranking with Large Language Models}
\author{
Guo Chen\textsuperscript{1,3,{\color{yuangreen}*}} \quad Wenhan Liu\textsuperscript{2} \quad Reed Li\textsuperscript{3} \quad Qianwen Ma\textsuperscript{3} \quad Yu Lu\textsuperscript{3,{\color{yuangreen}\dag}}\\[6pt]
Haibo Shi\textsuperscript{3} \quad Bingbing Xu\textsuperscript{4} \quad Junjie Huang\textsuperscript{1,{\color{yuangreen}\dag}}\\[6pt]
\textsuperscript{1}Southwest University \quad \textsuperscript{2}Renmin University of China \\
\textsuperscript{3}Yuanbao Team, Tencent \quad \textsuperscript{4}Institute of Computing Technology, Chinese Academy of Sciences \\
}
\newcommand{\paperrunningtitle}{GistRank: Gist-Compressed Listwise Ranking with Large Language Models}
\newlength{\barwidth}
  \makeatletter
  \newcommand\blfootnote[1]{%
    \begingroup
    \renewcommand\thefootnote{}%
    \renewcommand{\@makefntext}[1]{%
      \parindent=0pt\relax
      \noindent ##1%
    }%
    \footnote{#1}%
    \addtocounter{footnote}{-1}%
    \endgroup
  }
  \makeatother
\begin{document}

\maketitle
\blfootnote{\textsuperscript{{\color{yuangreen}*}}Work done during internship at Tencent Yuanbao. \textsuperscript{{\color{yuangreen}\dag}}Corresponding author}

\begin{abstract}
Large language models (LLMs) have become a powerful tool for listwise relevance ranking, which can improve web search by comparing retrieved passages jointly. However, processing every candidate verbatim and generating a permutation autoregressively can make inference costly, every candidate must be serialized verbatim into the prompt and the input therefore grows linearly with both the number and the length of the passages. While sliding-window ranking handles context limits through repeated re-encoding of overlapping blocks, hard compression techniques sacrifice the fine-grained details required for accurate discriminative judgment. To better understand and mitigate this trade-off, we conduct a comprehensive study of soft compression for LLM-based listwise ranking. Our findings reveal that a vanilla gist representation is ill-suited to ranking for two reasons: strictly causal encoding weakens how thoroughly each gist token absorbs its passage, and the representation is never trained to expose the relative-preference structure that separates closely competing candidates. To this end, we propose \ours, a three-stage training framework that equips a large language model with gist-compressed passage representations by learning faithful gist tokens under a semi-bidirectional attention scheme, instilling discriminative comparison through pairwise training, and adapting the model to the listwise objective. \ours~collapses each passage into only $K{=}4$ gist tokens, this gives 14$\times$ speedups over one-shot full-text ranking while reaching 59.17 average nDCG@10 over TREC DL19/20 and five BEIR subsets within 0.89 of the strongest full-context baseline. Overall, \ours~provides an effective means to enhance efficiency for searches in real-world environments.
\end{abstract}
\thispagestyle{firstpage}

\section{Introduction}

\begin{figure}[t]
    \centering
    \includegraphics[width=\linewidth]{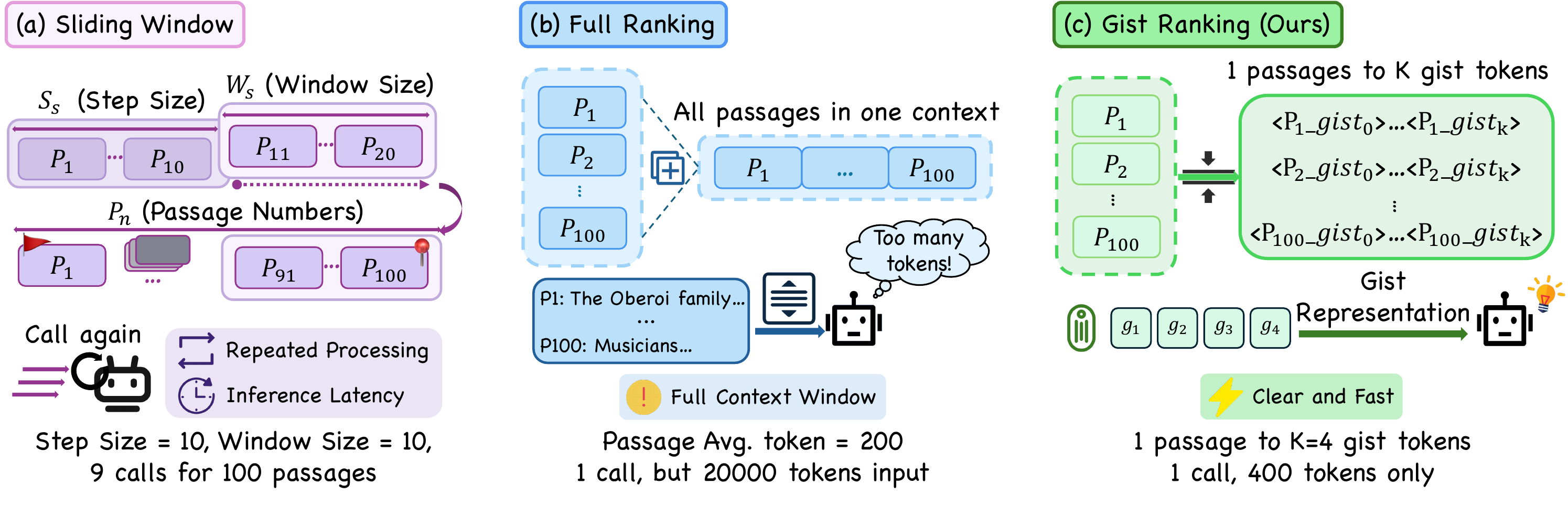}
    \caption{Comparison of existing listwise ranking strategies: Sliding Window, Full Ranking and \ours, corresponding to panels (a), (b), and (c), respectively.}
    \label{fig:fig1}
\end{figure}

Large language models have rapidly become a powerful tool for relevance ranking, where a single model is prompted with a query alongside an ordered list of candidate passages and asked to directly emit the permutation that best reflects their relevance~\citep{sun2023chatgpt,ma2023zero,liu2026training}. This listwise formulation lets the model reason jointly over all candidates and exploit cross-passage comparisons that pointwise and pairwise scorers cannot capture, and it has proven so effective that the capability of proprietary models has been distilled into compact open-source rankers that now serve as strong and reproducible baselines~\citep{pradeep2023rankvicuna,pradeep2023rankzephyr}. Driven by these successes, listwise ranking with large language models has quickly become a central paradigm in modern retrieval pipelines.
However, in web search and retrieval-augmented applications, the response must also be produced within an inference budget. A conventional listwise reranker serializes the candidate passages into the ranking prompt. Its input grows with both the number and length of passages, even when the entire list fits inside the model's context window~\citep{liu2024lost}. A long candidate list, or passages that are individually long, quickly  inflates inference latency, degrading both ranking quality and responsiveness. As shown in \figref{fig:fig1}, the prevailing remedy is to process the list through a sliding window that reranks overlapping blocks and propagates the ordering across passes~\citep{sun2023chatgpt,liu2025sliding}, but this comes at the cost of repeatedly re-encoding overlapping content. These costs motivate reducing the information consumed by the ranking stage, while explicitly accounting for any work needed to prepare that information.

A promising solution is to compress each passage before ranking, for example, extracting a summary or embedding the passage~\citep{feng2026sumrank, liu2025leveraging}. Besides, a family of soft-compression methods encodes long inputs into a handful of trainable gist tokens that retain the conditioning effect of the original text while drastically shortening the sequence~\citep{mu2023learning,ge2023context}. We argue that such compression is a natural fit for improving inference efficiency in ranking tasks, as it could replace each verbatim passage with a compact set of gist tokens. Prior work has in fact measured gist and KV-level compression on reranking and reported a large gap to full attention~\citep{deng2025silver}, but has treated that gap as an intrinsic cost of compression. We argue it is instead a feature of the compression objective: relevance judgment hinges on fine-grained, query-conditioned distinctions among closely competing candidates, whereas objectives optimised for reconstruction or generative conditioning are never asked to preserve those distinctions. In ranking, the compressed states must expose relative preference across candidates, not merely preserve a useful account of each passage. Bridging this gap therefore requires adapting the gist representation so that it preserves the discriminative signals on which ranking depends.

We propose \ours, which equips a large language model with compressed gist tokes passage representations for efficiency listwise ranking through a three-stage training procedure. In the first stage, we learn faithful gist representations: fixed-length documents are followed by a set of gist tokens that are in turn concatenated with the query, with the query trailing the gist tokens. The model is trained to predict the subsequent tokens conditioned solely on the gist, while granting the gist tokens richer bidirectional access to their source context. In the second stage, with the gist tokens visible, but the original documents hidden, we train these representations on real-world search engine interaction data with pairwise objectives, instilling the discriminative capability required to distinguish competing candidates for the purpose of ranking. In the third stage, we continue training the model directly on labeled ranking data, so that it learns to leverage the gist tokens to produce the listwise permutation both efficiently and accurately.

We evaluate both relevance and inference cost rather than treating a shorter prompt as proof of a faster complete pipeline. Over TREC DL19/20 and five BEIR subsets, \ours~obtains 59.17 mean nDCG@10. \ours~has total end-to-end P50 latencies of 1,625\,ms on DL19 and 1,634\,ms on DL20. Relative to the one-shot full-text baseline these correspond to 14.05$\times$ and 13.93$\times$ speedups. The measurement includes compression, KV assembly, and ranking. Direct stage measurements show that the compact rank forward accounts for only about 7\% of the model time: most work remains in candidate compression. This confirms that compressed segments can significantly reduce overhead during the inference phase. These results highlight the potential of gist-compressed ranking for scalable search applications.

Our main contributions are threefold:
\begin{itemize}
    \item We articulate the mismatch between gist objectives for compact conditioning and the relative-preference information required for listwise relevance ranking, and use it to motivate ranking-aware training.
    \item We propose \ours, a three-stage training method that first learns compressed gist representations, then continues training on real-world search engine data to enhance discriminative ability, and finally adapts gist tokens to efficient listwise ranking.
    \item We evaluate both ranking effectiveness and end-to-end latency on TREC DL19/20 and five BEIR subsets, demonstrating substantially lower latency than the controlled full-text baselines while maintaining competitive average nDCG@10. 
\end{itemize}

\section{Related Work}
In this section, we position \ours~at the intersection of listwise ranking and soft context compression, focusing on how discriminatively trained gist tokens can improve ranking efficiency while preserving effectiveness.

\subsection{Listwise Ranking}
Ranking a candidate set as a whole rather than scoring documents in isolation has long been recognized as a principled formulation, and the earliest learning-to-rank approaches directly optimized list-level objectives that account for the relative ordering among all candidates~\citep{cao2007learning,xia2008listwise,burges2010ranknet}. With the advent of pretrained language models, this paradigm was reframed as estimating relevance from deep contextualized interactions between a query and each document, first through cross-encoders that score pairs pointwise and then through sequence-to-sequence formulations~\citep{nogueira2019passage,nogueira2020document,zhuang2024beyond}. The emergence of large language models pushed this further by feeding an entire ordered list of passages into a single prompt and instructing the model to emit a permutation, which proved remarkably effective even in zero-shot settings and was subsequently distilled into smaller open models for practical deployment~\citep{reddy2024first,qin2024large}. Subsequent methods have taken both accuracy and efficiency into account—for instance, sliding-window-based approaches\citep{sun2023chatgpt}, as well as those utilizing reinforcement learning or collaborative ranking~\citep{liu2025reasonrank,liu2025coranking,liu2026demorank}. In real-world retrieval, deploying listwise rankers requires both strong ranking performance and manageable inference costs. Yet existing token-level listwise methods have largely overlooked gist tokens trained with a discriminative ranking objective as a way to compress passage representations while retaining ranking-critical information.

\subsection{Soft Context Compression}
Using soft context compression for model inputs has become an effective way to reduce the computational overhead of model inference. A line of work has sought to shorten the context consumed by a language model by encoding long inputs into a handful of trainable soft tokens, where the seminal approach learns dedicated gist tokens that compress an entire instruction into a compact prefix while preserving its conditioning effect on subsequent generation~\citep{mu2023learning}. Building on this idea, later efforts scaled the compression to arbitrary contexts by training autoencoding objectives that reconstruct the original text from a few memory slots, or by recursively summarizing long documents into accumulated soft prompts that extend the effective context length~\citep{chevalier2023adapting,ge2023context,deng2025silver}. More recent studies have pushed the compression ratio to extreme regimes\citep{petrov2025long,tarasov2025sentence, mao2026forget}, and integrated such condensed representations into retrieval-augmented pipelines to reduce the cost of conditioning on retrieved evidence~\citep{li2025500xcompressor,cheng2024xrag,lu2025turborag}. Furthermore, research has already extended pre-sorting compression to the token level in an effort to improve efficiency\citep{zhi2026compress, lu2026comprank}. However, these works are primarily optimized for tasks such as representation compression or generative guidance, while capabilities regarding discriminative comparison tasks have been overlooked. There has no research repurposed gist-style soft tokens as compact passage representations for ranking.

\section{Gist Compression for Ranking}
\subsection{Task Formulation}
\label{sec:task}
Given a user query $q$ and a list of candidate passages $P=[p_1,\dots,p_N]$ returned by a first-stage retriever, listwise ranking jointly compares the candidates and produces a permutation $\hat{\pi}$ that orders them by their relevance to $q$,
\begin{equation}
\hat{\pi} = f_{\mathrm{rank}}(q,P) \in \mathfrak{S}_N,
\label{eq:task}
\end{equation}
where $\mathfrak{S}_N$ is the set of all permutations of $\{1,\dots,N\}$, and $\hat{\pi}_k$ identifies the passage placed at rank $k$, with the aim of placing more relevant passages at earlier ranks. Let $x(\cdot)$ denote the token sequence of its argument. A full-text listwise ranker serializes all candidate passages into a ranking prompt. Its passage-content token count is
\begin{equation}
L_{\mathrm{full,passage}}
= \sum_{j=1}^{N}|x(p_j)|,
\label{eq:cost-full}
\end{equation}
which grows with both the number of candidates and their average token length. Our goal is to preserve joint comparison while reducing the passage content consumed by the ranking phase.

\subsection{Gist Compression}
\label{sec:gist}
For a query--passage pair $(q,p_j)$, the model first runs a compression phase over the serialized query and passage followed by $K$ gist tokens. The dense gist representation is defined as
\begin{equation}
\begin{split}
G^{(q,j)}
= \mathrm{Enc}_\theta\Big(
  [\,x(q);\,x(p_j);\, & \langle g_0\rangle,\dots,\langle g_{K-1}\rangle\,];\\
  & M_{\text{gist}}
\Big)\Big|_{\text{gist}}
\in \mathbb{R}^{K\times D}.
\end{split}
\label{eq:gist-enc}
\end{equation}
where $M_{\text{gist}}$ is the attention mask that forces the ranking-relevant information to pass through the gist positions. We further denote by $C^{(q,j)}$ the corresponding layer-wise KV-cache entries at these gist positions. The dense block $G^{(q,j)}$ is used for scoring, while $C^{(q,j)}$ is used as the compressed state consumed by the rank-phase forward pass.

With $K$ gist positions per passage, the candidate-content length in the rank phase becomes
\begin{equation}
L_{\mathrm{gist,passage}}=NK.
\end{equation}
These expressions describe the candidate-content length; query tokens, formatting tokens, and rank slots contribute additional positions to the complete rank prompt. Candidate compression is performed independently for each query--passage pair before joint ranking.

\subsection{Motivation for \ours}
\label{sec:analysis}
A gist trained only to predict text is encouraged to retain information sufficient for that prediction. Relevance ranking asks a different question: which of several candidate passages better satisfies this query? Passages may both contain terms that make the query predictable while differing in the specific evidence needed to order them. For example, a query about the cause of a condition may be predictable from passages that mention the condition, even if only one passage supplies the requested causal evidence. The representation must expose the distinction that affects their relative order, not just the shared topic.

\section{Supervised Fine-tuning for \ours}
\begin{figure}[t]
    \centering
    \includegraphics[width=\linewidth]{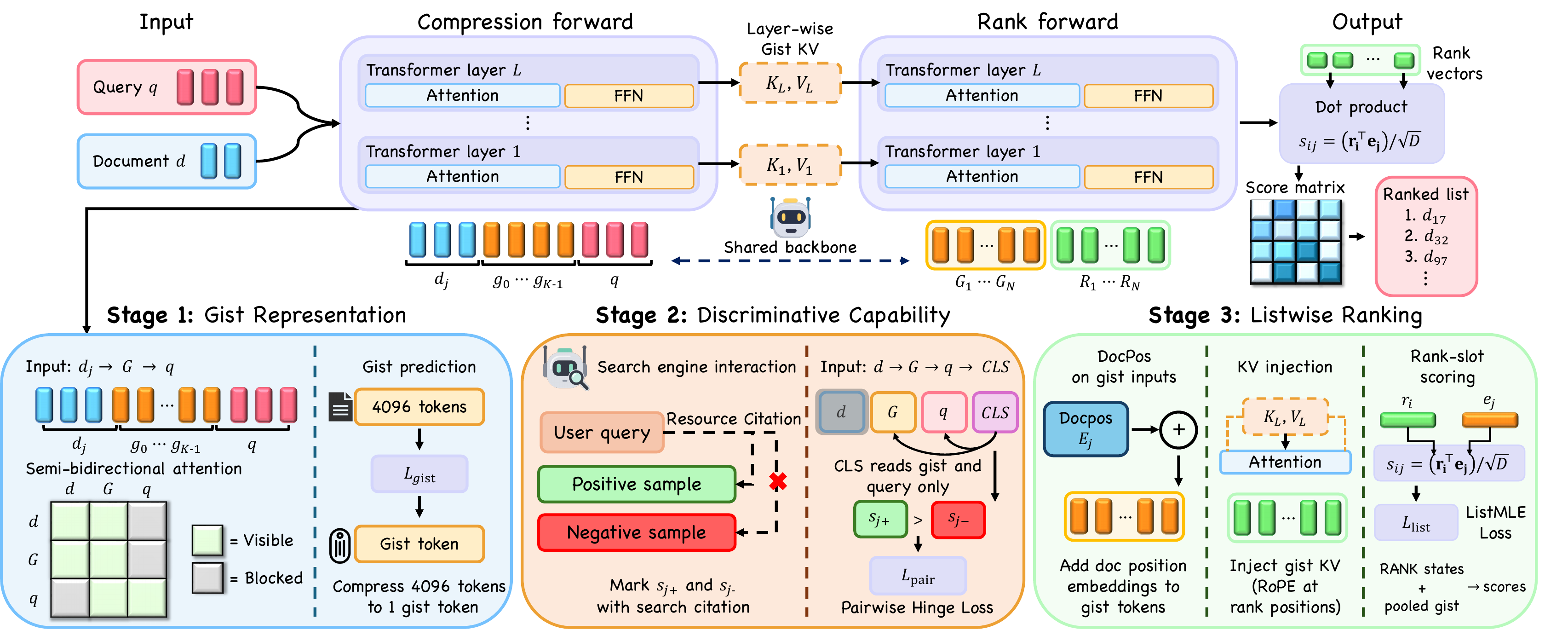}
    \caption{Overview of the training workflow. Stage 1 trains the ability to compress gist tokens. Stage 2 involves continued training to enhance discriminative capabilities using search data. Finally, Stage 3 aligns the model with listwise ranking objectives.}
    \label{fig:workflow}
    \vspace{-10pt}
\end{figure}

As shown in \figref{fig:workflow}, \ours~trains the same backbone in three stages: (i) query-predictive gist learning, (ii) pairwise preference training on search interactions, and (iii) listwise ranking adaptation. The gist token learned in earlier stages is reused throughout.
\subsection{Stage 1: Gist Representation}
\label{sec:stage1}

\paragraph{Stage 1: learn the gist interface.}
Stage~1 learns gist representations by predicting the query associated with a passage, following the sentence-compression formulation of \citet{tarasov2025sentence}. We use approximately 30 billion training tokens from search-engine queries and returned documents, packed into 4,096-token sequences. Observed query--passage associations provide supervision without additional relevance or preference labels.

The input is arranged as
\begin{equation}
[\,x(p_j);\langle g_0\rangle,\ldots,\langle g_{K-1}\rangle;x(q)\,].
\end{equation}
Under the semi-bidirectional mask, passage tokens attend within the entire passage, and gist tokens attend to all passage and gist tokens. Query tokens attend only to the gist states and preceding query tokens; passage and gist tokens cannot attend to the target query. Thus, passage information reaches query prediction through the gist tokens.

Let $G_{\mathrm{rec}}^{(j)}$ denote the passage gist states and $q_t$ the $t$-th query token. We minimize
\begin{equation}
\mathcal{L}_{\mathrm{gist}}
=-\frac{1}{|x(q)|}\sum_{t=1}^{|x(q)|}
\log p_\theta\!\left(q_t\mid q_{<t},G_{\mathrm{rec}}^{(j)}\right),
\label{eq:stage1}
\end{equation}
with loss computed only over non-padding query tokens, excluding passage and gist positions.

\subsection{Stage 2: Discriminative Capability}
\label{sec:stage2}
Stage~2 uses citation feedback from search-engine interactions as implicit preference supervision. For each query, the search engine returns results ordered by its ranking model. A result cited in the interaction is treated as $p^+$, and a returned but uncited result as $p^-$. We use approximately 15.36 million queries, each with one cited--uncited pair assigned labels 1 and 0. The texts come from the existing search data; citation feedback supplies the preference labels, no additional information is introduced.

For each query--passage pair $(q,p_j)$, the input is arranged as:
\begin{equation*}
[\,x(p_j);\,g_0,\ldots,g_{K-1};\,x(\text{query prompt}(q));\,\mathrm{CLS}\,]
\end{equation*}
The gist positions encode the passage before the query prompt is introduced. Under the attention mask, neither the query-prompt positions nor the final $\mathrm{CLS}$ position can attend directly to raw passage tokens, while $\mathrm{CLS}$ can attend to both the gist block and the query prompt. Consequently, the gist block serves as a compact document representation through which passage information reaches $\mathrm{CLS}$. Pairwise supervision encourages the query-conditioned $\mathrm{CLS}$ state to distinguish the cited result from the uncited one, and encourages the gist representation
to retain evidence useful for that distinction. 

Let $h_{\mathrm{cls}}^{(q,j)}$ denote the final hidden state at $\mathrm{CLS}$. A lightweight scoring head assigns 
\begin{equation}
s_j = \mathbf{w}_{\mathrm{pair}}^\top
h_{\mathrm{cls}}^{(q,j)} + b_{\mathrm{pair}}.
\end{equation}

Let $\mathcal{M}$ be the set of valid cited--uncited pairs within the same query in a training batch. We optimize a pairwise hinge loss:
\begin{equation}
\mathcal{L}_{\mathrm{pair}}
=
\frac{1}{|\mathcal{M}|+\epsilon}
\sum_{(j_+,j_-)\in\mathcal{M}}
\big[1-(s_{j_+}-s_{j_-})\big]_+.
\label{eq:stage2}
\end{equation}
Here, $[z]_+=\max(z,0)$, the margin is fixed to 1, and $\epsilon=10^{-9}$. A batch without a valid cited--uncited pair contributes zero pairwise loss.

\subsection{Stage 3: Listwise Ranking}
\label{sec:stage3}

Stage~3 adapts the learned gist representations to joint listwise ranking using the GPT-4o-annotated candidate orderings provided by \citet{liu2025sliding}.

\textbf{(i) Compression phase.}
We independently encode each query--passage pair as:
\begin{equation*}
[\,x(q);\,x(p_j);\,g_0,\ldots,g_{K-1}\,].
\end{equation*}
Unlike Stage~2, the query now precedes the gist tokens, so their states can encode query-conditioned passage information. Before compression, we add a learned document-position embedding to each gist-token input embedding:
\begin{equation}
\widetilde{\mathbf{u}}_{j,k}
=
\mathbf{u}(g_k)+\mathbf{E}^{\mathrm{doc}}_{j-1},
\qquad k=0,\ldots,K-1,
\label{eq:docpos}
\end{equation}
where $j$ is the candidate's position in the input list and the embedding table is indexed from zero. From this forward pass, we retain both the layer-wise gist keys and values and the final normalized gist hidden states.

\textbf{(ii) Rank phase.}
We assemble the candidates' gist slots into a single rank prompt followed by \textsc{rank} slots. At each transformer layer, the keys and values are replaced with those produced by the corresponding gist tokens; the keys use the slots' positions in the rank prompt for RoPE. The rank forward uses standard causal and padding masks. Let $\mathbf{r}_i$ be the final hidden state of the $i$-th \textsc{rank} slot, and let $\mathbf{e}_j=K^{-1}\sum_{k=0}^{K-1} \mathbf{h}^{\mathrm{comp}}_{j,k}$ be the mean of candidate $j$'s final normalized gist states. We score candidate $j$ at output position $i$ by
\begin{equation}
s_{ij}
=
\frac{\mathbf{r}_i^\top\mathbf{e}_j}{\sqrt{D}},
\label{eq:score}
\end{equation}
where $D$ is the hidden dimension.

In Stage~2, the gist represents the passage before seeing the query, so the query-conditioned $\mathrm{CLS}$ state performs the discriminative comparison. In Stage~3, the gist itself is query-conditioned; mean pooling yields one candidate vector shared across output positions, while the \textsc{rank} states capture listwise context.

For $N'$ valid candidates, let $\pi(i)$ identify the candidate at teacher rank $i$. We optimize a rank-slot formulation of ListMLE~\cite{xia2008listwise}:
\begin{equation}
\mathcal{L}_{\mathrm{list}}
=
-\frac{1}{N'}
\sum_{i=1}^{N'}
\log
\frac{\exp(s_{i,\pi(i)})}
{\sum_{j\in\mathcal{A}_i}\exp(s_{ij})},
\label{eq:listmle}
\end{equation}
Here, $\mathcal{A}_i=\{\pi(i),\ldots,\pi(N')\}$ contains the remaining teacher candidates. Gradients reach compression through pooled gist states and injected layer-wise keys and values. At inference, each rank slot selects the highest-scoring unselected candidate. Padded candidates are excluded from loss and selection.

\section{Experiment}
\subsection{Experimental Setup}
\subsubsection{Evaluation Datasets.}
We utilized the TREC DL 2019~\citep{craswell2020overviewtrec2019deep} and TREC DL 2020~\citep{craswell2021overviewtrec2020deep} datasets for our evaluation experiments, and also employed various datasets from the BEIR benchmark~\citep{thakur2021beir} for performance comparisons. Specifically, we selected five publicly available subsets from the BEIR collection (including Covid, DBPedia, SciFact, NFCorpus and Touche) to conduct evaluations.
\subsubsection{Baselines.}
For the primary efficiency experiments, we used Qwen3-4B~\citep{yang2025qwen3} as the base model and performed Supervised Fine-Tuning (SFT) on the same training set using two approaches: full-context ranking and ranking based on a summary extracted via BM25 for each passage. Each SFT model was implemented in both full-context and sliding-window variants, serving as direct benchmarks for efficiency. Regarding performance experiments, in addition to the basic BM25\cite{robertson2009probabilistic} baseline, we included the following models for comparison spanning pointwise, sliding-window, full-context, and embedding-based approaches: monoBERT (340M)~\citep{nogueira2019passage}, monoT5 (220M)~\citep{nogueira2020document}, RankT5 (3B)~\citep{zhuang2024beyond}, RankVicuna (7B)~\citep{pradeep2023rankvicuna}, RankZephyr (7B)~\citep{pradeep2023rankzephyr}, FullRank\textsubscript{100} (7B)~\citep{liu2025sliding}, and PE-Rank\textsubscript{BM25} (7B)~\citep{liu2025leveraging}. MonoBERT, monoT5, RankT5, PE-Rank and FullRank were trained on annotated query-passage pairs from the MS MARCO training set, whereas RankVicuna and RankZephyr were derived from ranking lists generated by GPT-4.
\subsubsection{Metrics.}
We evaluate our method from two perspectives: efficiency and retrieval quality.
For efficiency, we report three metrics:
\begin{itemize}
    \item \textbf{P50 latency (ms/query)} is the median end-to-end wall-clock time per query, reflecting the responsiveness perceived by users in online serving.
    \item \textbf{TFLOPs/query} is the number of floating-point operations per query, expressed in trillions, indicating how effectively a method utilizes the available compute.
    \item \textbf{Relative latency} is the ratio of a baseline's P50 latency to that of \ours~under the same evaluation environment.
\end{itemize}

For effectiveness, we adopt \textbf{nDCG@10}, which jointly assesses the relevance and the ranking order of the top-10 retrieved results, rewarding methods that place highly relevant documents near the top of the list.

\subsubsection{Implementation Details.}
We adopt Qwen3-4B-Base as the backbone to study efficient listwise reranking with a compact open-source language model. The backbone provides a 32k context window. For Stage~3, we reuse the MS MARCO-based candidate lists and GPT-4o-annotated teacher permutations provided by~\citet{liu2025sliding}, using a single training epoch by default. The controlled full-text, summary, and gist variants use the same Stage~3 training examples and teacher permutations. We compress each passage into 4 gist tokens, and adopt a listwise ranking window of size $100$. The model is optimized with a micro batch size of $1$ and a global batch size of $8$, using a learning rate of $5\times10^{-6}$. Training is conducted on a single node with 8$\times$H20 GPUs using the DeepSpeed framework~\citep{rasley2020deepspeed} and bfloat16 mixed precision. All evaluations, including both effectiveness and latency measurements, are performed on a single H20 GPU.

\begin{table}[t]
\centering
\small
\resizebox{\textwidth}{!}{%
\begin{tabular}{llrrrrrr}
\toprule
& & \multicolumn{3}{c}{DL19} & \multicolumn{3}{c}{DL20} \\
\cmidrule(lr){3-5}\cmidrule(lr){6-8}
Model & Strategy & P50 (ms) & TFLOPs/q & Relative latency & P50 (ms) & TFLOPs/q & Relative latency \\
\midrule
Qwen3-SFT-Full & Sliding window & 43,304 & 132.34 & 26.65$\times$ & 32,093 & 132.24 & 19.64$\times$ \\
Qwen3-SFT-Full & One-shot full list & 22,827 & 71.44 & 14.05$\times$ & 22,760 & 71.62 & 13.93$\times$ \\
Qwen3-SFT-Summary & Sliding window & 42,169 & 56.95 & 25.95$\times$ & 30,595 & 55.31 & 18.72$\times$ \\
Qwen3-SFT-Summary & One-shot full list & 18,312 & 28.35 & 11.27$\times$ & 18,305 & 28.22 & 11.20$\times$ \\
\ours  & Gist & \textbf{1,625} & 158.66 (85.05) & 1.00$\times$ & \textbf{1,634} & 160.16 (86.12) & 1.00$\times$ \\
\bottomrule
\end{tabular}}
\caption{Efficiency comparison on DL19 and DL20. Relative latency is each method's P50 divided by that of \ours. Parentheses give FLOPs excluding padding. Latencies cover the complete reranking pipeline.}
\label{table:efficiency}
\end{table}

\begin{table}[t]
\centering
\resizebox{\linewidth}{!}{
\begin{tabular}{cc|cccccccc}
\toprule
\makebox[2.4cm][c]{\textbf{Models}} &
\makebox[2.4cm][c]{\textbf{Strategies}} &
\textbf{DL19} &
\textbf{DL20} &
\textbf{Covid} &
\textbf{DBPedia} &
\textbf{SciFact} &
\textbf{NFCorpus} &
\textbf{Touch\'e} &
\textbf{Avg.} \\
\midrule
\textbf{BM25}        & --        & 50.58 & 47.96 & 59.47 & 31.80 & 67.89 & 33.75 & \underline{44.22} & 47.95 \\
\textbf{monoBERT(340M)}    & Pointwise & 70.72 & 67.28 & 73.45 & 41.69 & 62.22 & 34.92 & 30.26 & 54.36 \\
\textbf{monoT5(220M)}      & Pointwise & 70.58 & 67.33 & 75.94 & 42.43 & 65.07 & 35.42 & \textbf{46.98} & 57.68 \\
\textbf{RankT5(3B)}      & Pointwise & 72.29 & 69.14 & 79.03 & \textbf{44.85} & 74.60 & \underline{38.19} & 34.66 & 58.97 \\
\textbf{RankVicuna(7B)}  & Sliding   & 67.72 & 65.98 & 79.19 & \underline{44.51} & 70.67 & 34.51 & 33.00 & 56.51 \\
\textbf{RankZephyr(7B)}  & Sliding   & \underline{73.39} & 70.02 & \textbf{82.92} & 44.42 & 75.42 & 31.41 & 30.22 & 58.26 \\
\textbf{FullRank\textsubscript{100}(7B)} & Full & 72.55 & \textbf{71.29} & 82.24 & 43.54 & \textbf{77.04} & \textbf{39.14} & 34.63 & \textbf{60.06} \\
\textbf{PE-Rank\textsubscript{BM25}(7B)}     & Embedding    & 70.48 & 63.54 & 77.72 & 40.05 & 69.38 & 36.39 & 33.06 & 55.80 \\
\midrule
\multirow{2}{*}{\textbf{Qwen3-SFT-Full}}
    & Sliding & 72.83 & 69.47 & 77.67 & 42.18 & 74.68 & 37.50 & 30.24 & 57.80 \\
    & Full    & 72.58 & 68.30 & 76.52 & 41.70 & 73.34 & 36.17 & 31.57 & 57.17 \\
\midrule
\multirow{2}{*}{\textbf{Qwen3-SFT-Summary}}
    & Sliding & 64.25 & 58.59 & 67.18 & 36.09 & 70.08 & 33.33 & 31.11 & 51.52 \\
    & Full    & 64.45 & 61.14 & 69.11 & 35.60 & 70.34 & 33.58 & 32.19 & 52.34 \\
\midrule
\textbf{\ours} & Gist Token & \textbf{73.88} & \underline{70.82} & \underline{82.42} & 43.20 & \underline{75.46} & 36.27 & 32.15 & \underline{59.17} \\
\bottomrule
\end{tabular}
}
\caption{Effectiveness nDCG@10 results across seven datasets. \textbf{Bold} represents the best results, while \underline{underlined} denotes the second-best.}
\label{table:ndcg_results}
\end{table}

\subsection{Efficiency Performance}
In this section, we conduct efficiency experiments on DL19 and DL20 to verify that \ours~substantially accelerates listwise ranking, which is the central motivation of this work. Specifically, we rerank the top-100 passages retrieved by BM25 and compare three ranking regimes: the conventional sliding window (with a window size of 20 and a step size of 10), full ranking that consumes the entire list in a single pass, and our gist-token ranking. As controlled full-context baselines, we build two SFT variants on the same Qwen3-4B backbone as \ours: one that serializes every passage verbatim, and one that first condenses each passage into a BM25-extracted summary. We report end-to-end P50 latency, TFLOPs per query, and relative latency with respect to \ours. When calculating the TFLOPs for \ours~, we report both the computational cost associated with the padding process and the computational cost attributed to the actual padded tokens.

Table~\ref{table:efficiency} reports the corrected SFT efficiency measurements on DL19 and DL20. One-shot full-list ranking consistently achieves lower latency than sliding-window ranking. On DL19, processing the full list reduces P50 latency from 43{,}304\,ms to 22{,}827\,ms, and using BM25 passage summaries further lowers it to 18{,}312\,ms. The corresponding DL20 latencies are 32{,}093\,ms, 22{,}760\,ms, and 18{,}305\,ms. These results suggest that repeated processing of overlapping passages and input length both contribute to ranking latency. \ours~compresses the candidates into gist states and performs joint ranking in one non-autoregressive forward pass, with total P50 latencies of 1{,}625\,ms on DL19 and 1{,}634\,ms on DL20, including candidate compression and KV assembly. Relative to one-shot full-text ranking, these latencies give 14.05$\times$/13.93$\times$ speedups; relative to full-text sliding-window ranking, the factors are 26.65$\times$/19.64$\times$. The corresponding factors for summary-based one-shot and sliding-window ranking are 11.27$\times$/11.20$\times$ and 25.95$\times$/18.72$\times$, respectively. The gains reflect the complete inference procedure, including batched candidate compression and non-autoregressive ranking, rather than the isolated effect of shortening the rank prompt. Therefore, \ours~demonstrates a significant efficiency advantage throughout the entire inference process.

\subsection{Effectiveness Performance}

In this section, we evaluate the ranking effectiveness of \ours~. We rerank the top-100 passages retrieved by BM25 and report nDCG@10 on seven datasets, including the two TREC deep-learning sets DL19 and DL20 as well as five BEIR subsets covering diverse domains, namely TREC-COVID, DBPedia, SciFact, NFCorpus, and Touch\'e. We compare against a broad set of baselines that spans pointwise, sliding-window, full-context, and embedding-based models, and further include two Qwen3-4B variants trained under full-context and summary-based settings. Finally, we report the nDCG@10 results for each subset, as well as the average performance across the seven datasets.

The results are presented in Table~\ref{table:ndcg_results}. Despite compressing each passage into only a few gist tokens, \ours~achieves an average nDCG@10 of 59.17, ranking second among all evaluated methods and trailing the strongest baseline, FullRank$_{100}$, by 0.89 points. On the in-domain TREC datasets, \ours~achieves the highest score on DL19 at 73.88 and the second-highest score on DL20 at 70.82. These results suggest that compact gist representations retain relevance information needed for effective ranking. Compared with the Qwen3-SFT-Full baselines built on the same backbone, \ours~improves average nDCG@10 by 1.37 points over sliding-window ranking and 2.00 points over full-list ranking, demonstrating that its compact representations support competitive effectiveness even when compared with methods that retain the original passage text. In \secref{sec:ablation}, we establish a more direct performance reference by employing a full-text rank slot variant based on the uncompressed outputs from Stage 2. The comparison with Qwen3-SFT-Summary further highlights the effectiveness of learned soft compression: \ours~exceeds the summary-based baselines by 6.83--7.65 points on average. This gap suggests that learned gist representations preserve ranking-relevant information more effectively than BM25-extracted summaries. On the out-of-domain five BEIR datasets, performance varies across domains, but \ours~remains competitive on Covid and DBPedia, achieving scores of 82.42 and 43.20, respectively. Its overall average also exceeds those of the evaluated baselines, although individual methods retain advantages on particular datasets. These results indicate that \ours~can significantly reduce ranking latency while still maintaining robust overall ranking performance.
\begin{table}[t]
\centering
\resizebox{\textwidth}{!}{%
\begin{tabular}{llrrrrrrrr}
\toprule
Configuration & Candidate input & DL19 & DL20 & COVID & DBPedia & SciFact & NFCorpus & Touch\'e & Avg. \\
\midrule
Stage 1 $\rightarrow$ Stage 2 $\rightarrow$ Stage 3 (\ours) & Gist & 73.88 & 70.82 & 82.42 & 43.20 & 75.46 & 36.27 & 32.15 & 59.17 \\
Stage 1 $\rightarrow$ Stage 3 (w/o Stage 2) & Gist & 71.34 & 66.55 & 80.61 & 41.19 & 72.74 & 34.11 & 30.23 & 56.68 \\
\midrule
Matched full-text reference & Full text & 75.15 & 67.86 & 80.24 & 45.02 & 74.99 & 36.53 & 40.80 & 60.09 \\
\bottomrule
\end{tabular}}
\caption{Ablation results in nDCG@10 across seven datasets. The full-text reference shares the Stage~2 initialization, Stage~3 training data, hyperparameters, training steps, and rank-slot readout with \ours.}
\label{table:stage_ablation}
\end{table}

\begin{table}[t]
\centering
\setlength{\tabcolsep}{5pt}
\resizebox{0.95\textwidth}{!}{%
\begin{tabular}{lrrrrrrrr}
\toprule
\multirow{4}{*}{{\large Inference stage}}
& \multicolumn{4}{c}{DL19}
& \multicolumn{4}{c}{DL20} \\
\cmidrule(lr){2-5}\cmidrule(lr){6-9}
& \multicolumn{2}{c}{Latency}
& \multicolumn{2}{c}{Computation}
& \multicolumn{2}{c}{Latency}
& \multicolumn{2}{c}{Computation} \\
\cmidrule(lr){2-3}\cmidrule(lr){4-5}
\cmidrule(lr){6-7}\cmidrule(lr){8-9}
& Mean (ms) & Share & TFLOPs/q & Share
& Mean (ms) & Share & TFLOPs/q & Share \\
\midrule
Candidate compression
& 1,450.23 & 92.85\% & 149.09 & 93.97\%
& 1,463.74 & 92.89\% & 150.59 & 94.03\% \\
KV assembly
& 1.30 & 0.08\% & -- & --
& 1.32 & 0.08\% & -- & -- \\
Joint ranking + scoring
& 110.39 & 7.07\% & 9.56 & 6.03\%
& 110.69 & 7.02\% & 9.57 & 5.97\% \\
Total
& 1,561.93 & 100\% & 158.66 & 100\%
& 1,575.75 & 100\% & 160.16 & 100\% \\
\bottomrule
\end{tabular}%
}
\caption{Stage-wise inference cost of \ours~on DL19 and DL20. Latency is reported as the per-query mean. Computation include padding and exclude KV assembly.}
\label{table:stage_cost}
\end{table}

\begin{wraptable}{r}{0.48\textwidth}
\centering
\small
\captionsetup{font=small}
\resizebox{\linewidth}{!}{
\begin{tabular}{lrr}
\toprule
Input-order variant & DL19 & DL20 \\
\midrule
Shuffled order & 73.95 & 70.24 \\
Original order & 73.88 & 70.82 \\
\bottomrule
\end{tabular}
}
\caption{Input-order for \ours: nDCG@10 with shuffled and original candidate orders.}
\label{table:input_order}
\end{wraptable}
\subsection{Ablation Study}
\label{sec:ablation}

Table~\ref{table:stage_ablation} examines the contribution of Stage~2 and the effectiveness retained by gist compression. Since Stage~3 provides the listwise training required by the final ranking task, we retain it in both gist variants and ablate only Stage~2. Removing Stage~2 reduces average nDCG@10 from 59.17 to 56.68, a decrease of 2.49 points, with degradation across all seven datasets. The largest drop occurs on DL20, from 70.82 to 66.55, followed by SciFact, from 75.46 to 72.74. These consistent improvements suggest that learning to compress passages in Stage~1 does not fully capture the discriminative information needed for ranking, and that the citation-derived preference supervision in Stage~2 provides a useful initialization for subsequent listwise training. At the same time, the variant without Stage~2 remains competitive after Stage~3, indicating that Stage~2 strengthens ranking effectiveness rather than being a prerequisite for learning the ranking task.

The matched full-text reference shares the Stage~2 initialization, Stage~3 data, hyperparameters, training steps, and 100-rank-slot readout with \ours. Both compute the two-dimensional scores $s_{ij}=\mathbf{r}_i^\top\mathbf{e}_j/\sqrt{D}$ from \equref{eq:score}, optimize the same ListMLE objective, and use identical permutation selection. Only the candidate representation changes from explicit passage content to gist tokens. We use this as a full-text baseline under the same configuration, serving as no compression effectiveness reference, while ensuring that no context window overflow occurs. The full-text reference obtains 60.09 average nDCG@10 versus 59.17 for \ours, a difference of 0.92 points. This result shows that gist compression retains much of the effectiveness achievable by the same scoring design with access to explicit passage content.

Moreover, we evaluate input order as an inference-time variant by shuffling the candidate list before reranking. Table~\ref{table:input_order} compares this variant with the original input order on DL19 and DL20. Shuffling changes nDCG@10 from 73.88 to 73.95 on DL19 and from 70.82 to 70.24 on DL20. These small changes, with opposite directions across datasets, suggest that \ours~has limited sensitivity to the initial candidate order in this experiment. This is consistent with ranking decisions guided by query-conditioned candidate representations rather than dominated by the original retrieval order. This demonstrates that our method is robust to the ordering of input positions.
\subsection{Inference-cost accounting}
\label{sec:cost}
We analyze the inference cost of \ours~by separating candidate compression, KV assembly, and joint ranking. Latency measurements cover the complete reranking pipeline, from processing the query and candidate passages to producing the final ranked candidate list. Table~\ref{table:stage_cost} shows that candidate compression dominates inference latency, accounting for approximately 92.9\% of total time on both datasets. Its mean latency is 1,450.23\,ms on DL19 and 1,463.74\,ms on DL20, whereas joint ranking requires only 110.39\,ms and 110.69\,ms, respectively. KV assembly contributes less than 0.1\% of total time. Including all three stages, total P50 latency is 1,624.66\,ms on DL19 and 1,634.37\,ms on DL20. These results show that gist representations make listwise ranking inexpensive, with query-conditioned candidate encoding accounting for most of the remaining overhead.

The computation breakdown follows the same pattern. Candidate compression requires 149.09 and 150.59 TFLOPs/q on DL19 and DL20, accounting for approximately 94\% of total estimated operations. Joint ranking requires only 9.56 and 9.57 TFLOPs/q, bringing the totals to 158.66 and 160.16 TFLOPs/q. Although these totals exceed those of the one-shot baselines, \ours~achieves lower latency through batched candidate compression and joint ranking over compact gist states. The stage-wise results identify candidate compression as the main target for further latency improvements.

\subsection{Parameter Analysis}
In this section, we analyze the effect of the number of gist tokens $K$, which is used to compress each passage in Stage~3. This critical hyperparameter controls the number of inserted gist tokens, representing the trade-off between the compression rate and the amount of relevance signal preserved for each segment. To isolate the general representational capacity of gist tokens, we remove the explicit query $q$ from the Stage~3 ranking format and retrain the model while keeping all other settings fixed. This diagnostic evaluates whether the compressed passage states themselves remain useful for ranking. We vary $K \in \{1,2,4,8,16,32,128\}$ and report the average nDCG@10 across all seven evaluation datasets.

\begin{wrapfigure}{r}{0.48\textwidth}
    \centering
    \captionsetup{font=small}
    \includegraphics[width=\linewidth]{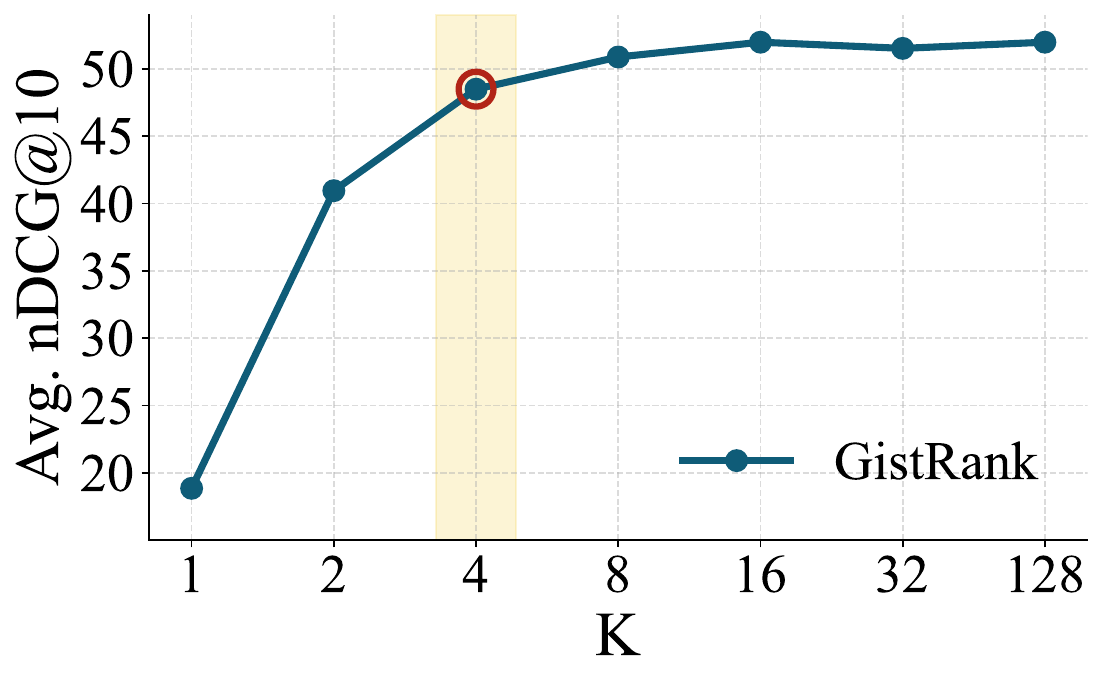}
    \caption{Parameter analysis across seven datasets. Performance rises sharply up to $K=4$ and then saturates, making $K=4$ a favorable efficiency--effectiveness trade-off.}
    \label{fig:k}
\end{wrapfigure}

The results are presented in Figure~\ref{fig:k}. Performance increases rapidly when $K$ grows from $1$ to $4$, showing that a small number of gist tokens is sufficient to recover useful passage signal. After $K=4$, however, the curve enters a clear saturation region: adding more gist tokens brings only marginal gains while increasing the context length and computation. This diminishing return suggests that larger gist representations provide limited additional benefit for ranking, while requiring the joint rank phase to process more candidate-token positions. We therefore view $K=4$ as the elbow point of the efficiency--effectiveness trade-off. It preserves sufficient ranking quality while keeping the representation compact, and we adopt $K=4$ as the default hyperparameter.

\subsection{Token Cost}
\begin{figure}[t]
\centering
\begin{tcolorbox}[
  width=\linewidth,
  colback=blue!5!gray!5,
  colframe=blue!30!gray,
  colbacktitle=blue!15!gray!15,
  boxrule=0.3mm, arc=2mm,
  left=6pt, right=6pt, top=5pt, bottom=5pt,
  title={Case Study: \emph{``causes of left ventricular hypertrophy''}},
  fonttitle=\bfseries\footnotesize\color{black},
  coltitle=black
]
\footnotesize
\textbf{(a) Full Context} \hfill \textbf{115 tokens}
\vspace{1pt}\hrule\vspace{3pt}
{\itshape Chamber Hypertrophy and Enlargement. In hypertrophy the heart muscle becomes thicker. $\cdots$ a congenital misalignment of cardiomyocytes, resulting in hypertrophy.}

\vspace{6pt}
\textbf{(b) BM25 Summary} \hfill \textbf{28 tokens}
\vspace{1pt}\hrule\vspace{3pt}
{\itshape Left ventricular hypertrophy results from an increase in left ventricular workload, e.g., during hypertension or aortic valve stenosis.}

\vspace{6pt}
\textbf{(c) \ours~} \hfill \textbf{4 tokens}
\vspace{1pt}\hrule\vspace{3pt}
\texttt{<|gist\_0|> <|gist\_1|> <|gist\_2|> <|gist\_3|>}
\end{tcolorbox}
\caption{A single DL19 candidate passage under the three ranking protocols. Full Context feeds the whole passage (115 tokens, abbreviated here), BM25 Summary keeps one 28-token sentence, and \ours~represents the passage with $K{=}4$ gist tokens.}
\vspace{-10pt}
\label{box:case_study}
\end{figure}

\begin{wraptable}{r}{0.48\textwidth}
\centering
\small
\captionsetup{font=small}
\resizebox{\linewidth}{!}{
\begin{tabular}{c|c|cc|c}
\toprule
\multirow{2}{*}{\textbf{Strategy}} &
\multirow{2}{*}{\textbf{Per-psg.}} &
\multicolumn{2}{c|}{\textbf{Prompt ($N{=}100$)}} &
\multirow{2}{*}{\textbf{Reduct.}} \\
& & \textbf{DL19} & \textbf{DL20} & \\
\midrule
\textbf{Full}       & 79 & 8{,}358 & 8{,}590 & 1.0$\times$ \\
\textbf{Summary}    & 26 & 3{,}015 & 2{,}884 & 2.8$\times$ \\
\textbf{Gist Token} & \textbf{4} & \textbf{1{,}253} & \textbf{1{,}254} & \textbf{6.8$\times$} \\
\bottomrule
\end{tabular}
}
\caption{Rank-stage prompt length on DL19 and DL20 with $N{=}100$ candidates. Representing each passage with four gist tokens reduces the full-list prompt to about 1.25k tokens, yielding a 6.7$\times$/6.9$\times$ reduction over full-context serialization.}
\label{table:token_cost}
\end{wraptable}

In this section, we quantify the rank-stage token cost to explain, at the token level, why \ours~is both faster and more predictable than the baselines. As shown in Table~\ref{table:token_cost}, we measure the rank-stage prompt length on DL19 and DL20 for $N{=}100$ passages. At the per-passage level, serializing a passage verbatim consumes about 79 tokens and a BM25 summary about 26, whereas \ours~represents each passage with only 4 gist tokens in the rank phase. This propagates to the full listwise rank input: over the $N{=}100$ candidates, the median full-text prompt spans more than 8{,}000 tokens on both DL19 and DL20 and the summary prompt around 3{,}000, while \ours~needs only about 1{,}250. This result shows that \ours~drastically reduces the context length of the listwise rank phase. A more distinctive property of \ours~is that every candidate contributes exactly $K$ gist slots regardless of its raw length.  This near-constant rank-phase length makes the downstream latency, the KV-cache footprint, and the batching throughput more predictable. 

Figure~\ref{box:case_study} makes this compression concrete by rendering one candidate passage under the three protocols. The full text keeps all 115 tokens, the summary retains only the single sentence that directly answers the query, and \ours~collapses the whole passage into 4 gist tokens. Although the summary happens to preserve the answering sentence, it discards the passage's discussion of alternative causes that a ranker may need to distinguish this candidate from its competitors; \ours~keeps the discriminative signal that Stage~2 is trained to preserve by a fixed number of gist tokens. By combining a fixed-size ranking representation with competitive ranking performance and low inference latency, \ours~offers predictable rank-stage costs and practical potential for deployment in search systems.

\section{Conclusion}
In this paper, we introduce \ours, a three-stage training framework for gist-compressed listwise ranking. The framework combines query-predictive gist learning, citation-derived pairwise supervision, and listwise adaptation to optimize compact passage representations for relevance ranking. Experiments on TREC DL19/20 and five BEIR subsets show that \ours~achieves competitive average nDCG@10. \ours~reduces reranking latency relative to autoregressive SFT baselines, while approaching the effectiveness of a full-text reference with the same rank-slot scoring procedure. These results highlight the potential of \ours~for efficient deployment in practical search systems. The inference-cost breakdown shows that candidate compression accounts for most of the remaining latency, motivating further optimization of this stage. Future work will investigate query-independent passage compression that allows gist states to be precomputed and reused across queries.

\clearpage
\bibliographystyle{main}
\bibliography{custom}

\end{document}